      [1999/12/01 v1.4c Il Nuovo Cimento]
\documentclass{cimento}

\usepackage{graphicx}

\newcommand{\be}{\begin{equation}} 
\newcommand{\ee}{\end{equation}} 
\newcommand{\ba}{\begin{eqnarray}} 
\newcommand{\ea}{\end{eqnarray}}

\title{Electroweak Corrections to t-channel single top production at the LHC}
\author{E.~Mirabella\from{mpi}}
\instlist{\inst{mpi} Max-Planck-Institut f\"ur Physik (Werner Heisenberg
  Institut) F\"ohringer 
Ring 6, D-80805 M\"unchen, Germany }

\begin{document}

\maketitle

\begin{abstract}
We describe the  computation of the $\mathcal{O}(\alpha^3)$ 
corrections for single top (and single anti-top) production
in the $t$-channel at hadron colliders. We show also 
The genuine one loop SUSY contribution 
to these processes. Such contributions are generally quite modest in the mSUGRA 
scenario. The experimental observables would therefore only  
practically depend, in this framework, on the CKM $Wtb$ coupling. 
\end{abstract}

\section{Introduction}
\label{Sec:Intro}
One of the major goals of the LHC is the study of the properties of the top
quark. In this respect single top production processes offer the unique
possibility of a direct measurement of the entry   
$V_{tb}$ of the CKM matrix allowing non trivial tests of the properties 
of this matrix in the Standard Model (SM)~\cite{SMsingle}. Moreover 
single top production processes allow for a test of the $V-A$ structure of the
charged current weak interaction of the top by looking at the polarization
of this quark~\cite{VAsingle}. Such processes can be interesting in the
hunting for physics beyond SM; indeed new physics can manifest itself either via loop 
effects,
or  inducing non SM weak interactions 
or introducing new single top production channels~\cite{BSMchan}. \\

\noindent
Within the Standard Model single tops can be produced via three
different modes. At the LHC the t-channel production mode will be not only the dominant 
one~\cite{TOPreview} but also the best measured: CMS 
studies~\cite{Pumplin:2002vw}  conclude that, with $10$ fb$^{-1}$ of  
integrated luminosity, the (mostly systematic)  
experimental uncertainty of the cross section is reduced below the ten percent 
level. \\

\noindent
Such an experimental precision  requests a similarly  
accurate theoretical prediction of the observables of the process.  
In order to  achieve it the complete NLO  
calculation is required. In the SM this has been done for the QCD component  
of the t-channel, resulting in a relatively small (few percent)  
effect~\cite{TCqcd}.  
The electroweak effects have been computed very recently at the 
complete one loop level within the SM and the MSSM~\cite{Beccaria:2006, Beccaria:2008}.
Such computation is the topic of this paper.\\ 

\noindent
In particular in sect.~\ref{Sec:EWcorrections} we 
briefly describe the structure of the Electroweak (EW) corrections to the
process of t-channel single (anti-)top production 
focusing on the partonic processes leading to such corrections. 
In sect.~\ref{Sec:SUSY} we discuss the SUSY corrections to these processes
evaluated within the MSSM. Numerical results for the EW corrections to t-channel
single (anti-)top production at the LHC are presented in
sect.~\ref{Sec:Numerics}, where the numerical impact of the SUSY corrections
is discussed as well. Sect.~\ref{Sec:Conclusions} summarizes our results.

\section{Electroweak Corrections}
\label{Sec:EWcorrections}
Electroweak corrections to single (anti-)top production in the t-channel are
of  $\mathcal{O}(\alpha^3)$  and in an obvious notation they can be written as:
\begin{eqnarray} 
\label{Eq:Main1l}
d \sigma_{t\mbox{\tiny-prod.}}(S)     &=& \sum_{\mbox{\tiny{(q, q')}}}  \int_{\tau_0}^1 d \tau   
 \frac{dL_{qb}}{d \tau}           \Big(  
                                 d \sigma^{\mbox{\tiny ew}}_{q b \to q' t}(s)          +  
                                 d \sigma^{\mbox{\tiny ew}}_{q b \to q' t
                                   \gamma}(s)    \Big), \\
d \sigma_{\bar{t}\mbox{\tiny-prod.}}(S) &=& \sum_{\mbox{\tiny{(q, q')}}}
 \int_{\tau_0}^1 d \tau \frac{dL_{\bar q \bar b}}{d \tau} \Big( 
                                 d \sigma^{\mbox{\tiny ew}}_{\bar q \bar b \to \bar q' \bar t}(s)          +  
                                 d \sigma^{\mbox{\tiny ew}}_{\bar q \bar b \to
                                 \bar q' \bar t \gamma}(s) \Big).     
\end{eqnarray} 
Where $(q,q')=(u,d),(c,s),(\bar d, \bar u), (\bar s, \bar d)$, while $\tau_0 = m^2_t / S$ and $s = \tau S$. The
differential luminosity is defined as:
\be
\label{Eq:Lumi}
\frac{dL_{i j}}{d\tau}(\tau)= \frac{1}{1+\delta_{ij}}~
\int_{\tau}^1 \frac{dx}{x} \left[ f_i(x)f_j\left(\frac{\tau}{x}\right)+ 
f_j(x)f_i\left(\frac{\tau}{x}\right)\right] ,
\ee
$f_{i}(x)$ being the momentum distribution of the parton 
$i$ in the proton. We perform our computation in Feynman gauge setting the CKM
matrix to unity. Due to CP invariance the unpolarized cross section of the
partonic process $\bar{q}\bar{b} \to \bar{q}' \bar{t}(\gamma)$ is equal to
that of the process
$q b \to q' t(\gamma)$,
so in the following we will analyse only the partonic processes contributing
to single top production. 

\subsection{Virtual Corrections}
\label{SSec:Virtual}
First class of corrections entering eq.~(\ref{Eq:Main1l}) are the virtual
corrections to the generic partonic process  $qb\to tq'$.\\
\noindent
The starting point is the cross section for the $ub\to td$ process;  
the $\mathcal{O}(\alpha^3)$ corrections to the (unpolarized) differential
cross section to this process read:
\begin{equation} 
d \sigma^{\mbox{\tiny ew}}_{ub\to td} =  \frac{dt}{64 \pi s^2} \sum_{\mbox{\tiny spin}}    
2  \mbox{Re} \{ \mathcal{M}^{0 ~*} \mathcal{M}^1  \}, 
\label{Eq:Upp}
\end{equation} 
where $\mathcal{M}^0$ is the tree level amplitude of the
partonic process $ub \to dt$  while
$\mathcal{M}^1$ describes the corresponding  EW  one loop amplitude. The Mandelstam 
variables are defined as:
\be 
s = (p_b+p_u)^2,~~~
t = (p_b-p_t)^2,~~~ 
u = (p_b-p_d)^2 .
\ee
The diagrams related to $\mathcal{M}^{1}$  have  been generated with 
the help of \verb|FeynArts|~\cite{FeynArts}, the algebraic reduction of the  
one loop integrals is performed with the help of \verb|FormCalc|~\cite{FormCalc} and 
the scalar one loop integrals are  numerically evaluated using \verb|LoopTools|~\cite{LoopTools}. 
We treat UV divergences using  dimensional reduction  
while IR singularities are parametrized giving a small mass
$m_\gamma$ to the photon. 
The masses of the light quarks  are used as regulators of the collinear 
singularities and are set to zero elsewhere.  \\ 
\noindent
UV finite predictions can be obtained  by  renormalizing the parameters and the wavefunctions 
appearing in $\mathcal{M}^{0}$. In our case we have to renormalize the wavefunction of the  
external quarks, the mass of the W boson, the weak mixing angle $\theta_W$
and the electric
charge. We use 
the on shell scheme decribed in ref.~\cite{DennerHab}.  
This scheme uses the fine structure constant evaluated in the Thomson limit 
as input parameter. In order to avoid large logarithms arising from the runnning  
of $\alpha$ to the electroweak scale $M_W$, we renormalize the fine structure
constant in the $G_\mu$ scheme {\it i.e.} we define $\alpha$ in terms of the 
the Fermi constant $G_\mu$:
\be
\alpha= \frac{\sqrt{2}}{\pi}G_{\mu}M_W^2\sin^2 \theta_W.
\label{Eq:alfaGM}
\ee 
We consistently change the definition 
of the renormalization constant of the fine structure constant following the 
guidelines of ref.~\cite{WjetProd,WjetProd2}. \\ 
\noindent
As pointed out in ref.~\cite{HollikGMU} in this scheme the 
leading logarithms arising from the running of $\alpha$ are resummed 
to all orders in perturbation theory and
absorbed in the definition of $\alpha$. Moreover, in the case of charged
current processes, 
the universal enhanced terms of the 
type $\alpha~m_t^2 / M_W^2$ are included as well.
\\
\noindent 
The unpolarized differential cross section for the process  $\bar d b \to \bar
u t$ can be obtained from 
that of the process $ub\to td$ by crossing: 
\begin{equation} 
d \sigma^{\mbox{\tiny ew}}_{\bar d b \to \bar u t} = \frac{dt}{64 \pi s^2} \sum_{\mbox{\tiny spin}}    
2  \mbox{Re} \{ \mathcal{M}^{0 ~*}(s\to u,~u\to s)  \mathcal{M}^1(s\to u,~u\to
s)  \}   .
\label{Eq:Dbar} 
\end{equation} 
The  differential cross sections of the processes involving
$c$ and $\bar{s}$ are, 
in the massless limit, equal to those quoted in eq.~(\ref{Eq:Upp}) and in eq.~(\ref{Eq:Dbar}), respectively.

\subsection{Real Corrections}
\label{SSec:Real}
Another class of $\mathcal{O}(\alpha^3)$ corrections entering
eq.~(\ref{Eq:Main1l})
are the tree level contributions to the  partonic processes of t-channel single top production associated with the 
emission of a  photon $qb\to tq' \gamma$. \\
\noindent
The unpolarized differential cross section of these processes has been  generated and squared using 
\verb|FeynArts| and  \verb|FormCalc|. 
According to the KLN theorem~\cite{KLN} IR singularities and the 
collinear singularities  
related to the final state radiation cancel in sufficiently inclusive 
observables while the collinear singularities related to initial state 
radiation  have to be absorbed into the Parton Distribution 
Functions (PDF). \\
\noindent
In order to 
handle with  these divergences   
we use two different procedures: the dipole subtraction method and 
the phase space slicing method. \\ 
In the subtraction approach one has to  add and subtract 
to the squared amplitude  
an auxiliary function wich matches pointwise the squared amplitude 
in the singular region and 
such that it can be analytically integrated 
over the photon phase space. Different functions fullfilling these
requirements are available in literature,  we 
use the function quoted in ref.~\cite{Dipole}. 
In this reference   
explicit expression for the subtraction function and for its  
analytical integration is obtained within mass regularization using   
the so called Dipole Formalism~\cite{DipoleQCD}. \\ 
\noindent
According to the phase space slicing approach  
the singular region of the phase 
space is excluded by introducing a cutoff on the energy of the 
photon and on the 
angle between the photon and  
the massless quarks. In the regular region the phase space 
integration can be performed  numerically while in the singular region  
it can be done  analytically in the eikonal approximation, provided that the
cutoffs are small enough. 
The form of the differential cross section in the  
singular region is universal and its explicit expression in the soft (collinear) region  can be  
found in ref.~\cite{DennerHab}~(\cite{WjetProd2}). \\
\noindent 
The two methods are in good  numerical agreement. 

\subsection{Mass Factorization}
As pointed out in sect.~\ref{SSec:Real},
$\mathcal{O}(\alpha^3)$ corrections to partonic cross sections contain
universal 
initial-state collinear singularities
that have to  be absorbed into the PDFs choosing a factorization scheme.
We use the $\overline{\mbox{MS}}$ factorization scheme
at the scale $\mu_F = m_t$.  \\
\noindent
Concerning 
the choice of the parton distributions set, 
we follow ref.~\cite{WjetProd}. The calculation of the full $\mathcal{O}(\alpha)$
corrections to any hadronic 
observable must include  
QED effects in the DGLAP evolution equations. Such effects are taken into
account in the 
MRST2004QED PDFs~\cite{Martin:2004dh} which are NLO QCD.  
Since our computation is leading order QCD and since the 
QED effects are known to be small~\cite{Roth:2004ti}, we use the LO set CTEQ6L.

\section{SUSY Corrections}
\label{Sec:SUSY}
As already mentioned in sect.~\ref{Sec:Intro}, accurate knowledge of the cross
section of single top production processes allows a precise 
determination of the $V_{tb}$ entry of the CKM matrix. Nevertheless some
non-standard physics could alter the prediction of the cross section biasing the
determination of this parameter. We calculate the impact of the one loop
corrections on t-channel single (anti-)top production within the MSSM. \\

\noindent
The one loop SUSY QCD corrections to t-channel single top production have
been computed at 
LHC in ref.~\cite{Zhang:2006cx}. 
We include these corrections, re-computing them from scratch.  
Following a standard procedure in SUSY QCD, we
treat UV divergences 
using dimensional regularization. 
Moreover in  this case we have to renormalize 
only the wavefunctions of the squarks since the other renormalization constants do not  
have $\mathcal{O}(\alpha_s)$ corrections. These corrections are IR and
collinear safe. \\  

\noindent
To obtain the genuine SUSY EW corrections one has to cope with the different
structure of the Higgs sector in the MSSM and in the SM. These 
corrections were obtained re-calculating the full
$\mathcal{O}(\alpha^3)$ corrections  and then subtracting the SM 
corrections computed setting the SM Higgs mass equal to the mass of the lightest 
MSSM Higgs boson. The computation of $\mathcal{O}(\alpha^3)$ corrections
within the MSSM was performed according to the procedure described in sect.~\ref{SSec:Virtual}

\section{Numerical results}
\label{Sec:Numerics}
We consider the numerical impact of the corrections described previously by  
looking at different observables. 

\subsection{EW Corrections within the SM}
\begin{figure}
\includegraphics[width=6.7cm]{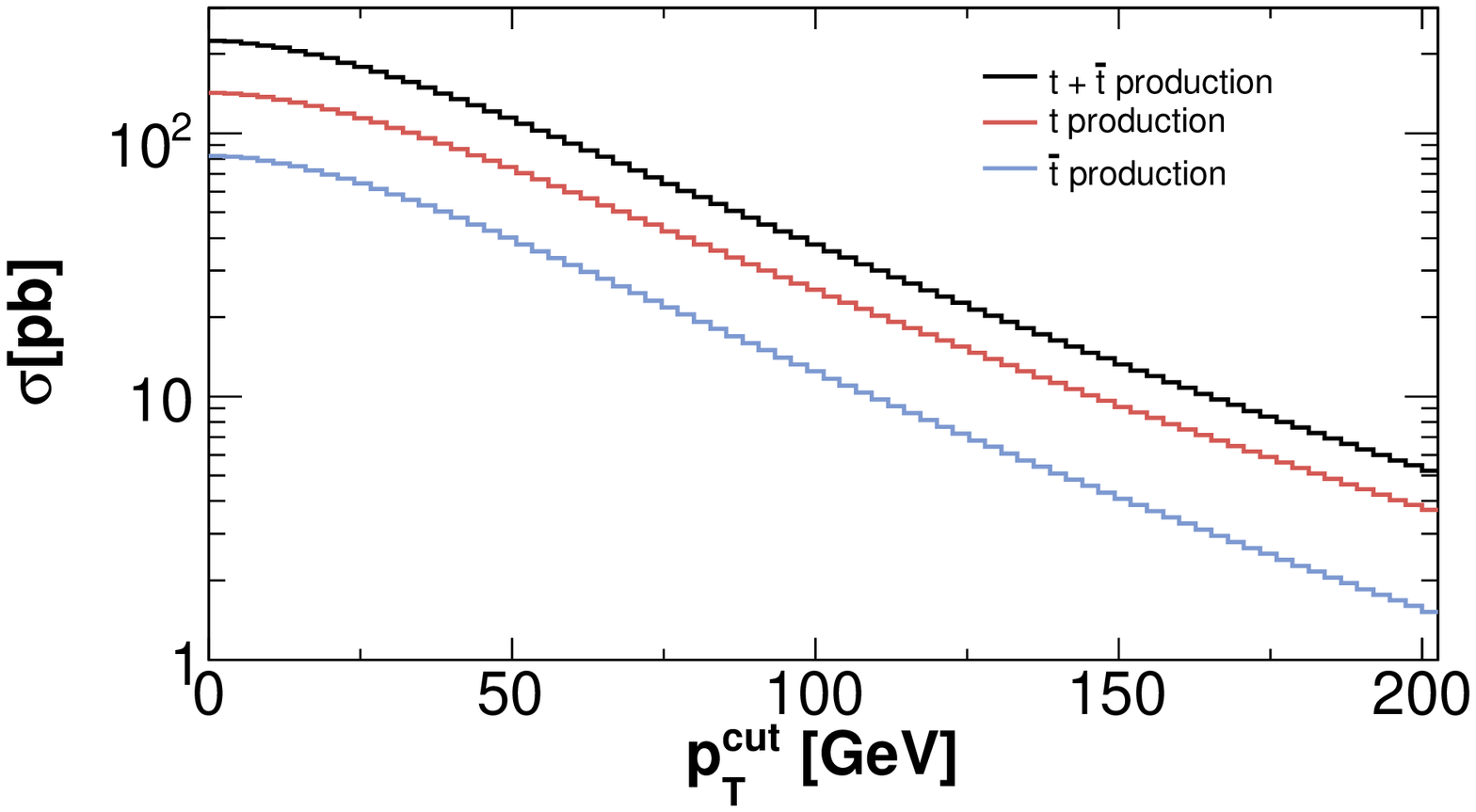}     
\includegraphics[width=6.7cm]{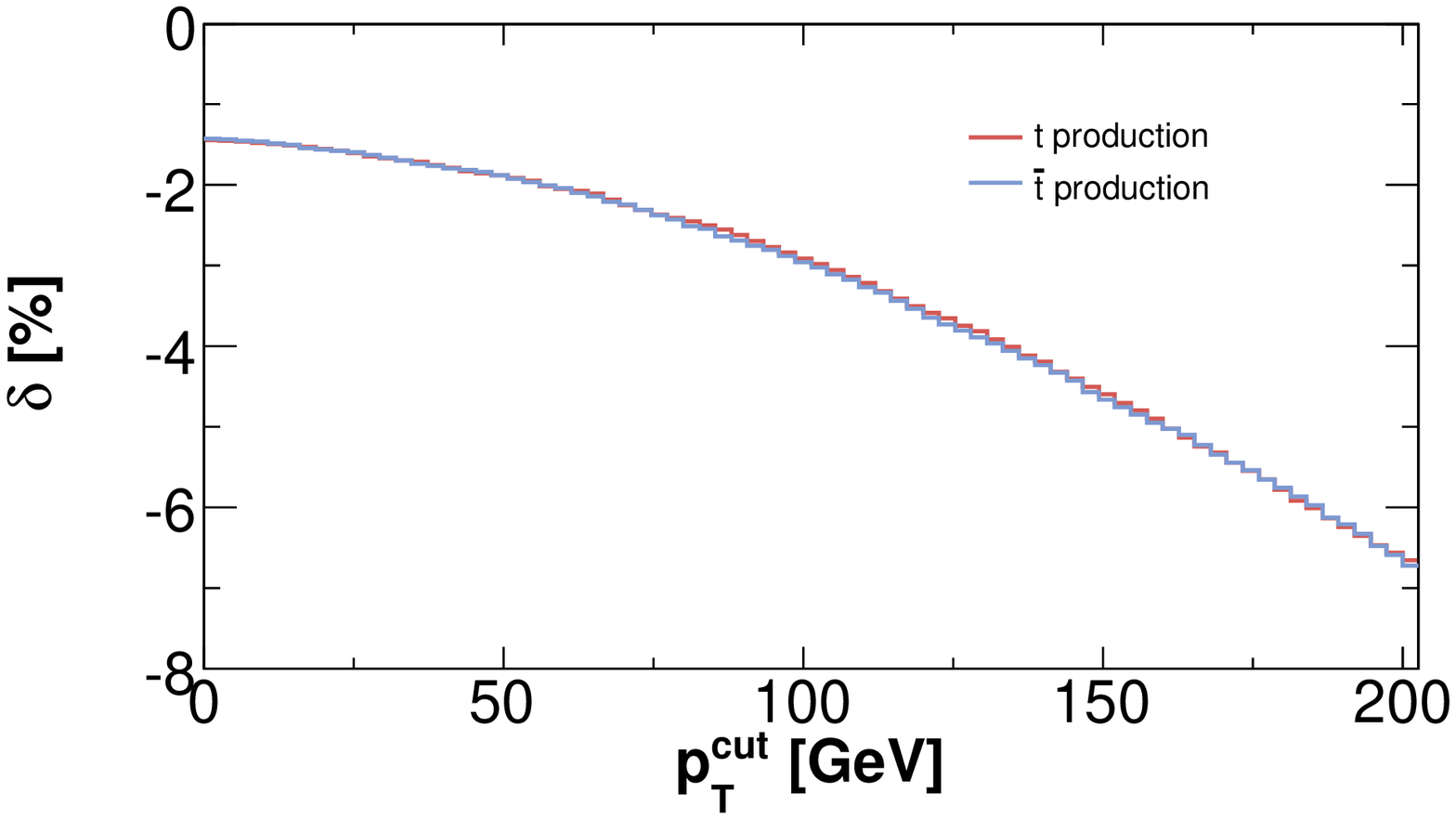}     
\caption{In the left panel we show the total cross section for single
 (anti-)top production 
as a function of a cut $p_T^{\mbox{\tiny cut}}$
on the transverse momentum of the (anti-)top. In the right panel the relative
contribution $\delta = [\sigma^{\mbox{\tiny NLO}}-\sigma^{\mbox{\tiny LO}}]/
\sigma^{\mbox{\tiny LO}}$
of the EW corrections to this observable is shown.}
\label{Fig:Fig01}
\end{figure}

\begin{figure}
\includegraphics[width=6.7cm]{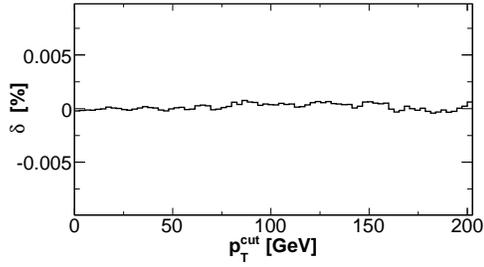}         
\caption{Relative contribution of the EW corrections to 
$R_{t\mbox{\tiny-prod.}/\bar{t}\mbox{\tiny-prod.}}$, defined as the ratio
between the total cross section for single top production and that for single
anti-top production.} 
\label{Fig:Fig03}
\end{figure}

\begin{figure}
\includegraphics[width=6.7cm]{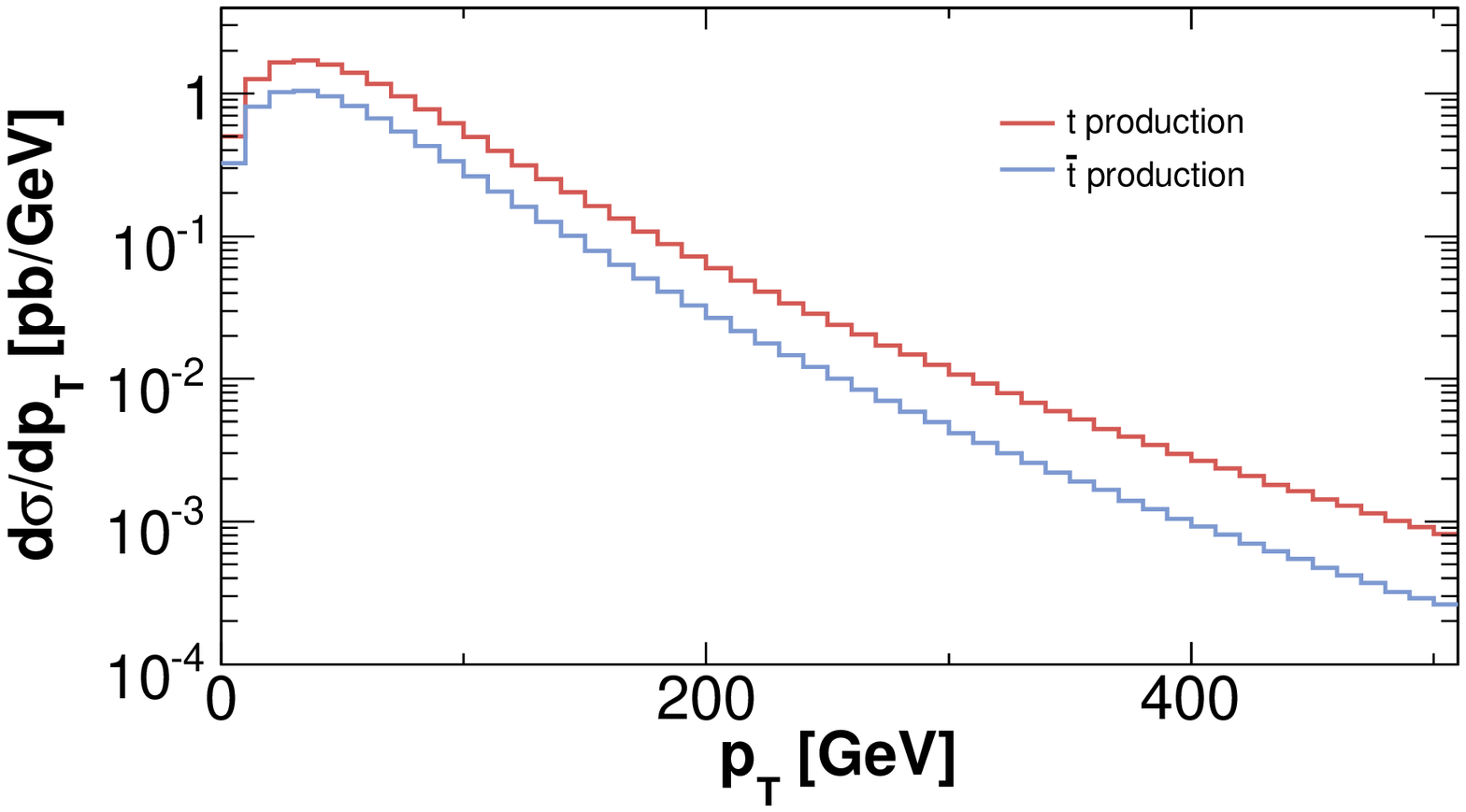}
\includegraphics[width=6.7cm]{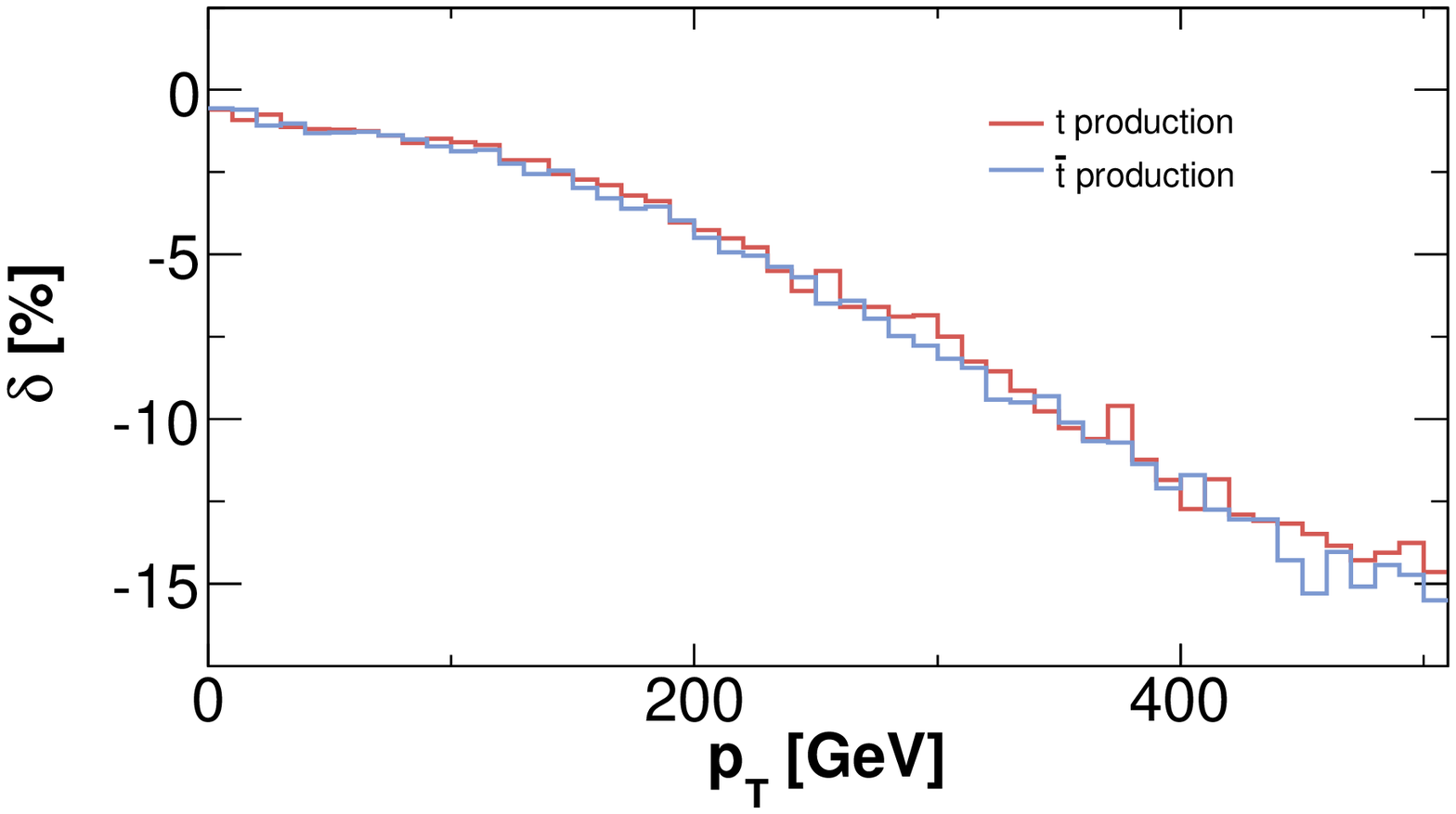}         
\caption{In the left panel the distribution of the transverse momentum
  of the (anti-)top  for single (anti-)top production is shown. In the right
  panel we show the relative
contribution of the EW corrections relative to the leading order result.}
\label{Fig:Fig04}
\end{figure}

In the left panel of fig.~\ref{Fig:Fig01} 
we show the NLO ({\it i.e.} LO +$\mathcal{O}(\alpha^3)$)
evaluation of the total cross section for 
single (anti-)top as a function of $p^{\mbox{\tiny cut}}_T$, a cut on the transverse momentum  
of the (anti-)top. As expected, at the LHC, single top production dominates 
over single anti-top production. Nevertheless, as can be inferred from the
right panel of
fig.~\ref{Fig:Fig01}, the relative contribution of the electroweak corrections
is similar in the two cases. Indeed in both cases EW corrections are negative
and become more important as the value of  $p_T^{\mbox{\tiny cut}}$
increases. EW corrections are well below $10~\%$ for any reasonable
value of $p_T^{\mbox{\tiny cut}}$. \\
\noindent
The similar behaviour of the EW
corrections in case of  single top and single anti-top production makes the quantity
\be
R_{t\mbox{\tiny-prod.}/\bar{t}\mbox{\tiny-prod.}} =
\frac{\sigma_{t\mbox{\tiny-prod.}}}{\sigma_{\bar{t}\mbox{\tiny-prod.}}},
\label{Eq:Ratio}
\ee
independent of the EW corrections (fig.~\ref{Fig:Fig03}). This is an
interesting feature since the value of the ratio~(\ref{Eq:Ratio})  
is used when looking for single top production processes at the
LHC~\cite{BSMchan}. \\
\noindent
In the left panel of fig.~\ref{Fig:Fig04} 
we show the NLO evaluation of
the 
transverse momentum ($p_T$)
distribution of the (anti-)top for the two production processes; while in 
right panel  the relative impact of the EW corrections is shown. 
This observable shows the same features as the previous one: in
particular the contribution of single top production is the leading one and
the relative contribution of the electroweak corrections is similar in both
single top and single anti-top production processes. Moreover EW corrections  are 
negative and their absolute value increases as the value of $p_T$ increases:
in particular they are larger  than  
 $10~\%$ in the region $p_T >300$~{\rm GeV}.

\subsection{SUSY Corrections}
\begin{figure}
\includegraphics[width=6.7cm]{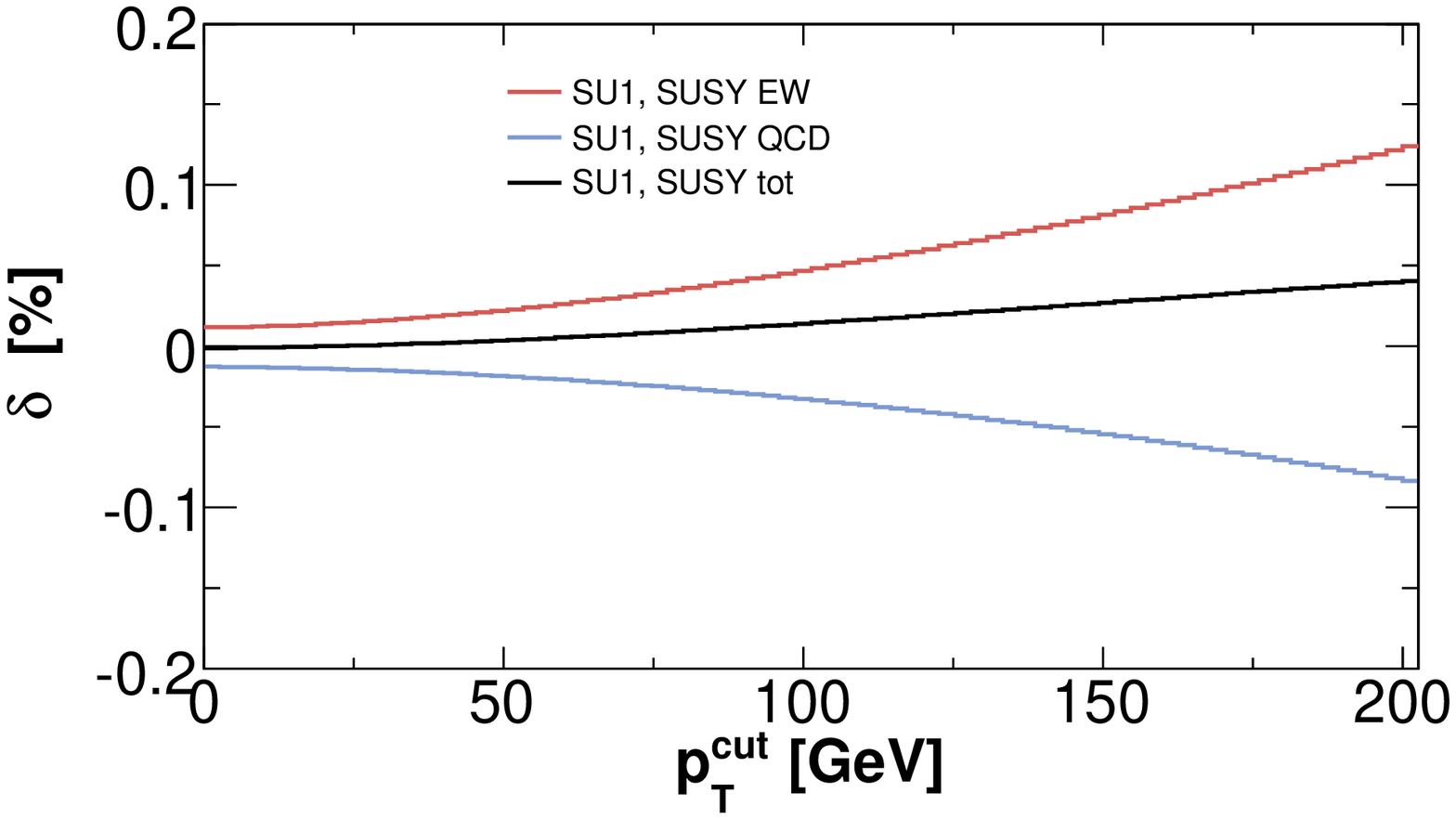}
\includegraphics[width=6.7cm]{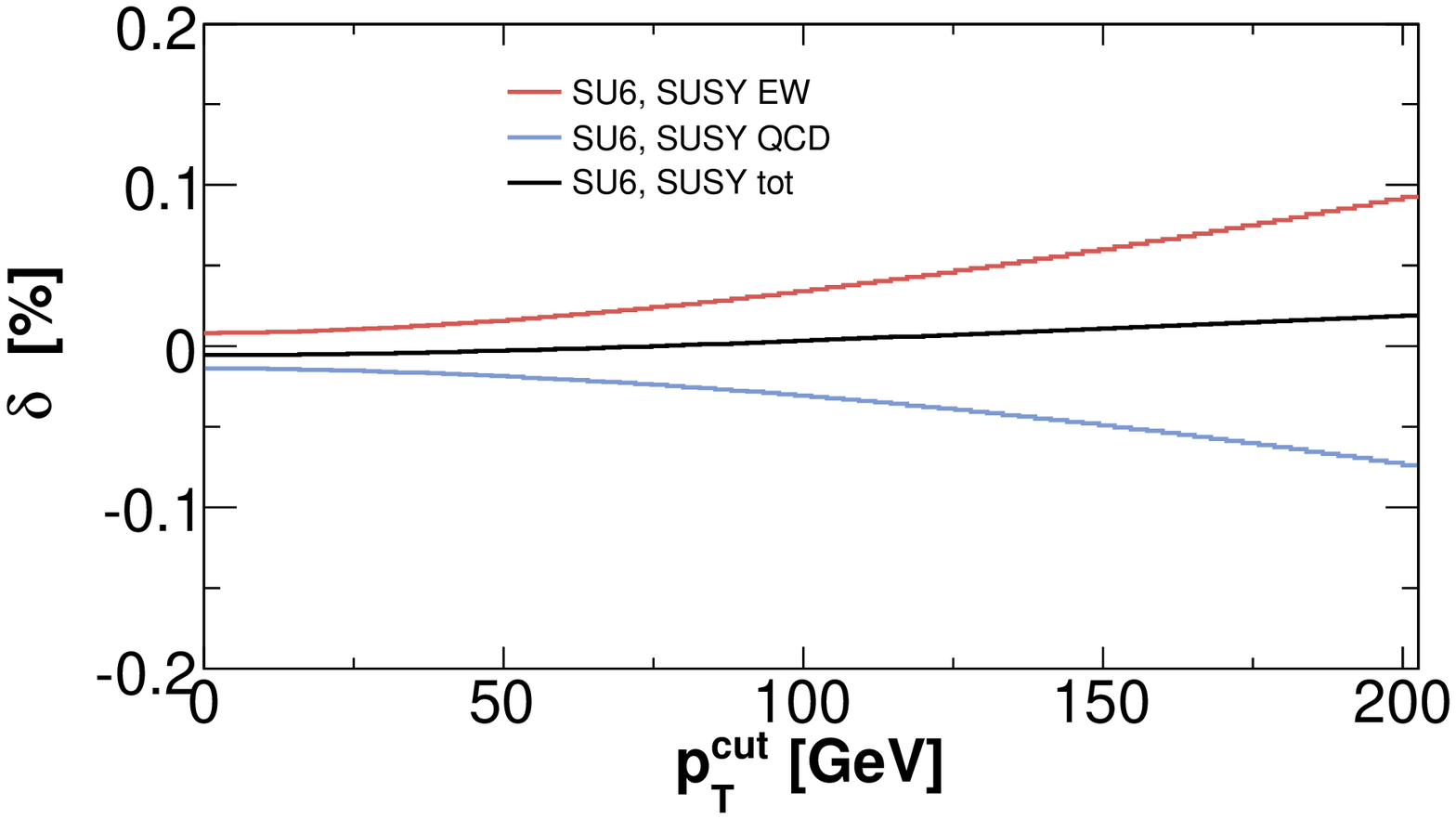}         
\caption{Contribution of the SUSY corrections to the total cross section of
single top and anti-top production as a function of a cut $p_T^{\mbox{\tiny
    cut}}$ on the transverse momentum of the (anti-)top. In both panels we show the 
contribution of the SUSY EW, of the  SUSY QCD corrections and of their sum relative to the leading order
contribution. In the left (right) panel such corrections are computed in the
 SU1 (SU6) scenario.}
\label{Fig:Fig06}
\end{figure}

We present the numerical results for the two representative
ATLAS DC2 mSUGRA benchmark points SU1 and SU6~\cite{DC2'}.
In fig.~\ref{Fig:Fig06}  
we show the relative contribution of the SUSY corrections to the total cross
section for single top and single anti-top production as a function of
$p_T^{\mbox{\tiny cut}}$  
in the case of the SU1 point (left panel) and SU6 point (right panel). In both
cases SUSY QCD and SUSY EW
corrections are tiny. Moreover they have different sign, so their sum is
further suppressed and in both cases  below $0.1~\%$.

\section{Conclusions}
\label{Sec:Conclusions}
We have computed the one loop EW corrections to the process of single
(anti-)top
production in the t-channel. 
The overall result is that the impact of these corrections on the total rate is small, 
of the order of a negative few percent. EW corrections can play a more
important role in the definition of the distributions. \\

\noindent
Moreover we have studied the impact of the one loop SUSY corrections within the MSSM. 
In the scenarios we have considered their impact is negligible and so their
eventual presence would not spoil the measurement of $V_{tb}$  performed 
within the SM.\\

\noindent
{\bf Acnowledgements}\\
\noindent
We would like to thank M.~Beccaria, C.M.~Carloni~Calame, G.~Macorini,
F.~Piccinini, F.M.~Renard and  C.~Verzegnassi for the pleasant collaboration 
in the work presented here. We gratefully acknowledge W.~Hollik for 
valuable suggestions.

\newpage

\end{document}